\documentclass[pra, twocolumn, amsfonts]{revtex4}
\usepackage{amsfonts}
\usepackage{amsmath}
\usepackage{bm}
\usepackage{graphicx}
\usepackage{epsf}
\usepackage{amssymb}
\usepackage{dsfont}

\begin{document}

\title{On the relation between Bell inequalities and nonlocal games}

\author{J. Silman}

\author{S. Machnes}

\author{N. Aharon}

\address{School of Physics and Astronomy, Raymond and Beverly Sackler
Faculty of Exact Sciences,
Tel-Aviv University, Tel-Aviv 69978, Israel}

\begin{abstract}{\bf
We investigate the relation between Bell inequalities and
nonlocal games by presenting a systematic method for their bilateral
conversion. In particular, we show that while to any nonlocal game there 
naturally corresponds a unique Bell inequality, the converse is not true. 
As an illustration of the method we present a number of nonlocal games that admit 
better odds when played using quantum resources.}
\end{abstract}

\date{January 8, 2007}

\maketitle

\section{Introduction}

Quantum mechanics admits stronger correlations between remote parties
than allowed by any (causal) classical theory \cite{Bell}. These
correlations, arising from the entanglement properties of the product
Hilbert space, can be used in various nonlocal games \cite{Cleve}
to improve on the maximum winning probability that obtains using only
classical correlations \cite{Brassard,Tsirelson,Vaidman GHZ,Vaidman necklace,Cabello,Aravind}.
As such, these examples also constitute proofs of the nonlocal nature
of quantum mechanics.

In this paper we explore the relation between nonlocal games, as defined by Cleve \emph{et al.} \cite{Cleve}, and Bell-type
\cite{Bell,CHSH} and GHZ-type \cite{GHZ} nonlocality proofs. In
particular, we show that to any member of a certain class of Bell
inequalities there ``naturally'' corresponds a nonlocal game.
To complement this we show the converse as well; any nonlocal game can be 
uniquely mapped to a Bell inequality in this certain class.

\section{Background}

\subsection{Nonlocal games}

As defined in \cite{Cleve}, a nonlocal game
is a cooperative task for a team of several remote players.
Every player is randomly assigned by a verifier an input according
to some joint probability distribution. Each then chooses one out of a set of possible outputs
and sends it to the verifier. The verifier consults a truth table dictating
for each combination of inputs, what combinations of outputs result
in a win. The players know the winning conditions, as well as the
joint probability distribution governing the assignment of combinations
of inputs, and may coordinate a joint strategy prior to receiving
them, but cannot communicate subsequently. A team making use of quantum
correlations (shared entanglement) is said to employ a 
``quantum
strategy'', whereas if not, is said to employ a ``classical strategy''.

\subsection{Bell inequalities}

In deterministic local hidden-variable theories all measurable quantities
are predetermined. Locality enters via the requirement that the results
of measurements carried out in any region are independent of what
type of measurements were, are or will be carried out, if at all,
in spacelike separated regions. In this light, let us consider two spacelike
separated parties $A$ and $B$ sharing an entangled state of a pair
of qubits. Suppose now that $A$ measures the spin component of his
qubit along some axis. The result that obtains must be independent
of the axis along which $B$ measures. This together with predeterminism
(realism) implies the following for two-spin measurement settings
per party \begin{equation}
a_{1}b_{1}+a_{1}b_{2}+a_{2}b_{1}-a_{2}b_{2}=\pm2\,,\label{algebraic}\end{equation}
where $a_{i}$ and $b_{i}$ denote the value of the spin component
of $A$'s qubit and $B$'s qubit along the axes ${\boldsymbol{\mathrm{n}}}_{A}^{(i)}$ and
${\boldsymbol{\mathrm{n}}}_{B}^{(i)}$, respectively ($i=1,\,2$). Averaging over many repeats
of the experiment, or what amounts the same thing, averaging over
the hidden-variable distribution, we obtain the CHSH inequality \begin{equation}
\left|\left\langle a_{1}b_{1}\right\rangle +\left\langle a_{1}b_{2}\right\rangle +\left\langle a_{2}b_{1}\right\rangle -\left\langle a_{2}b_{2}\right\rangle \right|\leq2\,,\label{CHSH}\end{equation}
or put differently, 
\begin{eqnarray}
 &  & 1\leq\left|P(a_{1}b_{1}=1)+P(a_{1}b_{2}=1)+P(a_{2}b_{1}=1)\right.\nonumber \\
 &  & \qquad\left.+P(a_{2}b_{2}=-1)\right|\leq3\,.\label{CHSH alt}\end{eqnarray}

Now eq. (\ref{algebraic}) is nothing more than algebraic relation
for two pairs of independent variables which assume the values $\pm1$.
Indeed, analogous relations exist for any number $n$ of such pairs
of variables \begin{equation}
\sum_{s}c_{s}\prod_{i=1}^{n}o_{i}^{(s_{i})}=-C,\,\dots,\, C\,,\label{algebraic generalize}\end{equation}
where $o_{i}^{(s_{i})}=\pm1$ and $s$ is an $n$-component vector
with $s_{i}=1,\,2$, the summation carried out over all possible vectors.
And give rise to a whole host of Bell inequalities for ``full''
correlation functions of dichotomic outcomes (i.e. Bell inequalities
that only consider the events $\prod_{i=1}^{n}o_{i}^{(s_{i})}=\pm1$
with $n$ the number of parties) \begin{equation}
\bigl|\sum_{s}c_{s}\bigl\langle \prod_{i=1}^{n}o_{i}^{(s_{i})}\bigr\rangle \bigr|\leq C\,,\label{CHSH generalize}\end{equation}
or alternately, as weighted sums \begin{equation}
S_{C}^{min}\leq\sum_{s}w_{s}P\bigl(O_s=\ell_{s}\bigr)\leq S_{C}^{max}\,,\qquad\ell_{s}=\pm1\,,\label{CHSH alt generalize}\end{equation}

The $c_{s}$ and $C$, and the $w_{s}$, $S_{C}^{min}$ and $S_{C}^{max}$
are of course related, and may easily be obtained
from one another. For example, suppose we are given any one of the inequalities
eq. (\ref{CHSH generalize}). Using the identities 
\begin{widetext}
\begin{equation}
\Bigl\langle \prod_{i=1}^{n}o_{i}^{(s_{i})}\Bigr\rangle =P\bigl(\prod_{i=1}^{n}o_{i}^{(s_{i})}=1\bigr)-P\bigl(\prod_{i=1}^{n}o_{i}^{(s_{i})}=-1\bigr)=2P\bigl(\prod_{i=1}^{n}o_{i}^{(s_{i})}=1\bigr)-1=1-2P\bigl(\prod_{i=1}^{n}o_{i}^{(s_{i})}=-1\bigr)\,,\label{identities}\end{equation}
\end{widetext} 
 to substitute $2P\bigl(\prod_{i=1}^{n}o_{i}^{(s_{i})}=1\bigr)-1$
and $1-2P\bigl(\prod_{i=1}^{n}o_{i}^{(s_{i})}=-1\bigr)$ for $\bigl\langle \prod_{i=1}^{n}o_{i}^{(s_{i})}\bigr\rangle $
whenever $c_{s}$ is positive or negative, respectively, and rearranging
terms, we obtain \begin{equation}
w_{s}=2|c_{s}|\,,\quad S_{C}^{min}=C-\sum_{s}|c_{s}|\,,\quad S_{C}^{max}=C+\sum_{s}|c_{s}|\,.\label{trans1}\end{equation}

Most generally, we may consider $n$ spacelike separated parties, where
the $i$-th party may measure $m_{i}$ different observables, corresponding
to different settings of their measurement device. Note that $m_{i}$
need not equal $m_{j}$. Moreover, the different measurement settings
employed by the same party need not have the same number of distinct
outcomes. Any Bell inequality pertaining to this system admits a weighted
average representation as follows 
\begin{eqnarray}
 &  & S_{C}^{min}\leq\sum_{M}w_{M}P(o_{1}^{(s_{1})}=\lambda_{1}^{(s_{1},\,r_{1})},\,\dots,\, o_{n}^{(s_{n})}=\lambda_{n}^{(s_{n},\,r_{n})})\nonumber \\
 &  & \qquad\leq\, S_{C}^{max}\,.\label{Bell general}\end{eqnarray}
Here, given that player $i$ employs the measurement setting $s_{i}$, $r_{i}$ labels the different possible outcomes.  $P(o_{1}^{(s_{1})}=\lambda_{1}^{(s_{1},\,r_{1})},\,\dots,\, o_{n}^{(s_{n})}=\lambda_{n}^{(s_{n},\,r_{n})})$
is the probability that party $1$ obtain the result $\lambda_{1}^{(s_{1},\,r_{1})}$
when employing the measurement setting $s_{1}$ ($s_{1}=1,\,\dots,\, m_{1}$), party $2$ obtain
the result $\lambda_{2}^{(s_{2},\,r_{2})}$
when employing the measurement setting $s_{2}$ ($s_{2}=1,\,\dots,\, m_{2}$), etc.
$M$ is a vector of ordered pairs  $(s_{i}, r_{i})$. Note that the summation is
carried out over all possible $M$, i.e. all possible measurement settings and outcomes.

\section{From Bell inequalities to nonlocal games}

Before presenting the method for converting Bell inequalities to nonlocal
games \emph{\'{a} la} Cleve \emph{et al.} \cite{Cleve}, we would
like to motivate it on an intuitive level. Roughly speaking, the equivalence
between the pair hinges on two key points of similarity: (i) In nonlocal
games each player must choose his output without knowing the input
assigned to any of the others. Similarly, in local hidden-variable
theories the value of a physical quantity measured in one region obtains
independently of which physical quantities were, are, or will be measured,
if at all, in spacelike separated regions. (ii) There are no classical
nonlocal game strategies that only allow for correct or preferable combinations
of outputs (unless the game is trivial). In a like manner, local
hidden-variable theories never saturate the algebraic limit of (nontrivial)
Bell inequalities.\\

Consider the family of Bell inequalities, eq. (\ref{CHSH generalize}). The
first step in converting these into games is to suitably reinterpret
the expectation values in this new context. To do so we present the concise Bell inequalities
- nonlocal games dictionary. In the {}``language'' of nonlocal games
$s_{i}$ denotes the input received by player $i$, while $o_{i}^{(s_{i})}$
represents his output. $\bigl\langle \prod_{i=1}^{n}o_{i}^{(s_{i})}\bigr\rangle $
is therefore the expectation value of the product of outputs given
the set of inputs $s$. The hidden-variable indicates the choice of
strategy, and the averaging is understood to be carried out with respect
to the different strategies employed \cite{foot1}.

Next we need to introduce joint probability distributions to govern
the assignment of inputs and truth tables. To this end, let us shift
our attention to the equivalent formulation of these Bell inequalities,
eq. (\ref{CHSH alt generalize}).
These relations hold for independent sets of dichotomic variables,
whether these variables describe physical quantities or outputs in
a nonlocal game. However, for the sums in these relations to make
sense in the context of nonlocal games, we have to give meaning to
the $w_{s}$. This is easily achieved by normalization, that is, we set the joint probability 
distribution for the inputs such that
\begin{equation}
\varrho_{s}=\frac{w_{s}}{\sum_{s}w_{s}}=\frac{|c_{s}|}{\sum_{s}|c_{s}|}\,.\label{prob distribution}\end{equation}
If we now construct the truth tables such that the games are considered
to have been won iff \begin{equation}
\prod_{i=1}^{n}o_{i}^{(s_{i})}=\ell_{s}\,\hat{=}\left\{ \begin{array}{cc}
+1 & c_{s}>0\\
\,-1 & c_{s}<0\end{array}\right.\,,\label{winning conditions}\end{equation}
where $o_{i}^{(s_{i})}=\pm1$ is the output of player $i$, the superscript
$s_{i}$ serving to denote its (possible) dependence on the input,
then the games' total winning probabilities are given by \cite{foot2}
\begin{equation}
P_{C}^{min}\leq\sum_{s}\varrho_{s}P\bigl(\prod_{i=1}^{n}o_{i}^{(s_{i})}=\ell_{s}\bigr)\leq P_{C}^{max}\,,\label{winning prob}\end{equation}
with \begin{equation}
P_{C}^{max/min}=\frac{S_{C}^{max/min}}{\sum_{s}w_{s}}=\frac{C\pm\sum_{s}|c_{s}|}{2\sum_{s}|c_{s}|}\,.\label{max classical}\end{equation}
the maximum and minimum classical total winning probabilities.\\

Using a quantum strategy the classical maximum total winning probability
can be surpassed. To do so the players must share a suitable entangled
state. Upon receiving his input, each player measures the spin component
of his qubit in a direction such that over many repetitions of the
game a maximal violation of the originating Bell inequality would
obtain. The maximum total winning probability is therefore given by
\begin{equation}
P_{Q}^{max}=\frac{S_{Q}^{max}}{\sum_{s}w_{s}}=\frac{Q+\sum_{s}|c_{s}|}{2\sum_{s}|c_{s}|}\,,\label{max quantum}\end{equation}
where $Q$ and $S_{Q}^{max}$ denote the upper bounds imposed by quantum
mechanics on the sums in eqs. (\ref{CHSH generalize}) and (\ref{CHSH alt generalize}),
respectively. This gives an advantage of \begin{equation}
\frac{Q-C}{2\sum_{s}|c_{s}|}=\frac{S_{Q}^{max}-S_{C}^{max}}{\sum_{s}w_{s}}\label{advantage}\end{equation}
to the optimal quantum strategy over the optimal classical one.\\

Bell inequalities for full correlation functions of dichotomic outcomes
are part of a larger class of Bell inequalities, which have in common
that in their weighted sum form, eq. (\ref{Bell general}),
nonvanishing coefficients, $w_{M}$, pertaining to the same measurement
settings are equal, and therefore independent of the outcome. Any member of this class
 can be converted into a nonlocal game. To see 
this we note that as weighted sums, 
eq. (\ref{Bell general}), these inequalities admit a simplified form
\begin{eqnarray}
& & S_{C}^{min}\leq\sum_{s}w_{s}\sum_{\mu}P(o_{1}^{(s_{1})}=\lambda_{1}^{(s_{1},\, \mu)},\,\dots,\, o_{n}^{(s_{n})}=\lambda_{n}^{(s_{n},\,\mu)})\nonumber \\ & & \qquad \leq\, S_{C}^{max}\,.\label{WIC}\end{eqnarray}
Here the summation over $\mu$ is carried out over different sets of outcomes,
which are not necessarily mutually exclusive. That is, 
$\lambda_{k}^{(s_{k},\,\nu\neq \mu)}$ may equal $\lambda_{k}^{(s_{k},\,\mu)}$. 
$P(o_{1}^{(s_{1})}=\lambda_{1}^{(s_{1},\, \mu)},\,\dots,\, o_{n}^{(s_{n})}=\lambda_{n}^{(s_{n},\,\mu)})$ is the probability that party $1$ obtain the result $\lambda_{1}^{(s_{1},\,\mu)}$
when employing the measurement setting $s_{1}$ ($s_{1}=1,\,\dots,\, m_{1}$), party $2$ obtain
the result $\lambda_{2}^{(s_{2},\,\mu)}$
when employing the measurement setting $s_{2}$ ($s_{2}=1,\,\dots,\, m_{2}$), etc. 
The construction of the joint probability distribution governing the
assignment of inputs and the truth table is analogous to that of
the full correlation functions case. The joint probability distribution
for the inputs is still obtained via eq. (\ref{prob distribution}).
However, the winning conditions can no longer be expressed by eq.
(\ref{winning conditions}). If
up to normalization eq. (\ref{WIC}) is to represent the game's total
winning probability, then given a combination of inputs $s$ the full
set of winning combinations of outputs must equal $\cup_{\mu}\bigl\{ \lambda_{1}^{(s_{1},\, \mu)},\,\dots,\,\lambda_{n}^{(s_{n},\, \mu)}\bigr\} $.
The game is then considered to have been won iff \begin{equation}
\bigl\{ o_{1}^{(s_{1})},\,\dots,\, o_{n}^{(s_{n})}\bigr\} \subseteq\cup_{\mu}\bigl\{ \lambda_{1}^{(s_{1},\, \mu)},\,\dots,\,\lambda_{n}^{(s_{n},\, \mu)}\bigr\} \,,\label{winning cond general}\end{equation}
where $\bigl\{ o_{1}^{(s_{1})},\,\dots,\, o_{n}^{(s_{n})}\bigr\}$
denotes some combination of outputs returned by the players.

\section{From nonlocal games to Bell inequalities}

When considering Bell inequalities for full correlation functions,
up to normalization, the conversion gives rise to a one to one mapping
between the coefficients of the Bell inequality, the $w_{s}$, and
those of the input frequencies of the nonlocal game, the $\varrho_{s}$.
See eq. (\ref{prob distribution}). (That the mapping is not one to
one between the $c_{s}$ and the $\varrho_{s}$, is merely due to the
fact that the Bell inequalities, eq. (\ref{CHSH generalize}), remain
unchanged if we flip the signs of all the $c_{s}$.) It is therefore
straightforward to invert this procedure and use it to obtain a Bell
inequality for full correlation functions from any nonlocal game with
dichotomic outputs.

This one to one character of the mapping carries over to the conversion
of any of the inequalities, eq. (\ref{WIC}). (See the last paragraph in the previous subsection.)
This leads to the conclusion that any nonlocal game can be converted
into a Bell inequality.

\section{Examples}

We now give two examples illustrating the application of our method.
In the first example we convert a family of Bell inequalities for
full correlation functions into a corresponding family of nonlocal
games. In the second example we illustrate the more general
case of non-full correlation Bell inequalities.

\subsection{Example I}

We consider the following family of two-qubit Bell inequalities for
$n\times n$ measurement settings introduced by Gisin \cite{Gisin}
\begin{equation}
\left(\begin{array}{cccccc}
\;\:\,1 &  & \cdots &  & \;\:\,1 & -1\\
 & \ddots &  & \;\:\,1 & -1\\
\vdots &  & \;\:\,1 & -1 &  & \vdots\\
 & \;\:\,1 & -1 &  & \ddots\\
\;\:\,1 & -1 &  & \ddots &  & -1\\
-1 &  & \cdots &  & -1 & -1\end{array}\right)\leq\left\{ \begin{array}{cc}
\frac{1}{2}n^{2} & even\, n\\
\\\frac{1}{2}(n^{2}+1) & odd\, n\end{array}\right.\,.\label{Gisin inequalities}\end{equation}
Here the matrix's dimension is $n\times n$ and its $(i,\, j)$-th
component denotes the coefficient of $\left\langle a_{i}b_{j}\right\rangle $.
For example, in this notation the CHSH inequality reads \begin{equation}
\left(\begin{array}{cc}
\;\:\,1 & -1\\
-1 & -1\end{array}\right)\leq2\,.\label{CHSH inequality}\end{equation}

Since all the $c_{ij}$ equal $\pm1$ it follows from eq. (\ref{prob distribution})
that the joint probability distribution should be set as uniform \begin{equation}
\varrho_{ij}=\frac{1}{n^{2}}\,.\label{Gisin distribution}\end{equation}
As for the truth table, eq. (\ref{winning conditions}) instructs
us to require that identical (opposite) outputs be returned given
the inputs $i$ and $j$ if the coefficient of $\left\langle a_{i}b_{j}\right\rangle $
is positive (negative). Examining eq. (\ref{Gisin inequalities}),
we see that the matrix's component are arranged such that \begin{equation}
c_{ij}=\left\{ \begin{array}{cc}
+1 & i+j\leq n\\
\,-1 & i+j>n\end{array}\right.\,.\label{Gisin coefficients}\end{equation}
The winning conditions therefore amount to the return of anticorrelated
outputs given inputs whose sum is greater than $n$, and correlated
otherwise. From eqs. (\ref{winning prob}) and (\ref{max classical})
we then have\\
 \begin{eqnarray}
 &  & \sum_{i+j\leq n}P(o_{A}^{(i)}=o_{B}^{(j)})+\sum_{i+j>n}P(o_{A}^{(i)}=-o_{B}^{(j)})\nonumber \\
 &  & \qquad\leq\left\{ \begin{array}{cc}
\frac{3}{4} & even\, n\\
\\\frac{3}{4}+\frac{1}{4n^{2}} & odd\, n\end{array}\right.\,.\label{Gisin probability}\end{eqnarray}
We see that as $n\rightarrow\infty$ the maximum total winning probability
converges to $75\%$. In this limit we can effect a transition to
the continuum. Introducing the variables \begin{equation}
\alpha=\lim_{n\rightarrow\infty}\frac{i}{n}\,,\qquad\beta=\lim_{n\rightarrow\infty}\frac{j}{n}\,,\label{continuous}\end{equation}
the game translates to the task of returning identical outputs whenever
$\alpha+\beta \leq 1$, and opposite outputs otherwise \cite{Nati}.

Higher probabilities can be reached using a quantum strategy. The
maximum obtains when the players share a singlet state, with one of
the players measuring at an angle of $\frac{i}{\pi n}$ spanning from,
say, the negative $x$-axis in the $xy$-plane, and the other at an angle
of $\frac{j}{\pi n}$ spanning from the negative $y$-axis in the
same plane. The dependence of the maximum on $n$ is given by 

\begin{equation}\label{g1 quant max}
P_{Q}^{max}=\frac{\cos(\frac{\pi}{2n})}{n\sin(\frac{\pi}{n})}+\frac{1}{2}\,,\qquad n\neq1\,,
\end{equation}
as is easily verified making use of eq. (\ref{max quantum}), with
$S_{Q}^{max}$ taken from \cite{Gisin}. Once again we see that as $n\rightarrow\infty$ the
maximum converges to a fixed value of $\simeq 81.8\%$.

\subsection{Example II}

We consider the following Bell inequality for three qutrits \cite{Acin}
\begin{widetext} \begin{eqnarray}
 &  & \left|P(o_{A}^{(1)}+o_{B}^{(1)}+o_{C}^{(1)}=0)-P(o_{A}^{(1)}+o_{B}^{(1)}+o_{C}^{(2)}=2)-P(o_{A}^{(1)}+o_{B}^{(2)}+o_{C}^{(1)}=2)+P(o_{A}^{(1)}+o_{B}^{(2)}+o_{C}^{(2)}=1)\right.\label{popescu}\\
 &  & \left.-P(o_{A}^{(2)}+o_{B}^{(1)}+o_{C}^{(1)}=2)+P(o_{A}^{(2)}+o_{B}^{(1)}+o_{C}^{(2)}=1)+P(o_{A}^{(2)}+o_{B}^{(2)}+o_{C}^{(1)}=1)+2P(o_{A}^{(2)}+o_{B}^{(2)}+o_{C}^{(2)}=0)\right|\leq3\,,\nonumber \end{eqnarray}
\end{widetext}
where $o_{A}^{(i)},\, o_{B}^{(j)}=1,\,0,\,-1$ and
all the equalities are evaluated modulus three. Substituting \begin{widetext}\begin{equation}
P(o_{A}^{(i)}+o_{B}^{(j)}+o_{C}^{(k)}=n)=1-P(o_{A}^{(i)}+o_{B}^{(j)}+o_{C}^{(k)}\neq n)=1-P(o_{A}^{(i)}+o_{B}^{(j)}+o_{C}^{(k)}=n+1)-P(o_{A}^{(i)}+o_{B}^{(j)}+o_{C}^{(k)}=n-1)\label{g2 identity}
\end{equation}\end{widetext}
for each of the probabilities in the second line of eq. (\ref{popescu})
we get 
\begin{equation}
0\leq \beta \leq 6 \, , \label{game 2}
\end{equation}
with
\begin{widetext}\begin{eqnarray}
 &  & \beta\hat{=} P(o_{A}^{(1)}+o_{B}^{(1)}+o_{C}^{(1)}=0)+P(o_{A}^{(1)}+o_{B}^{(1)}+o_{C}^{(2)}\neq2)+P(o_{A}^{(1)}+o_{B}^{(2)}+o_{C}^{(1)}\neq2)+P(o_{A}^{(1)}+o_{B}^{(2)}+o_{C}^{(2)}=1)
\nonumber\\
 &  & +P(o_{A}^{(2)}+o_{B}^{(1)}+o_{C}^{(1)}\neq2)+P(o_{A}^{(2)}+o_{B}^{(1)}+o_{C}^{(2)}=1)+P(o_{A}^{(2)}+o_{B}^{(2)}+o_{C}^{(1)}=1)+2P(o_{A}^{(2)}+o_{B}^{(2)}+o_{C}^{(2)}=0)\,.\nonumber\\
& & \;\label{beta} \end{eqnarray}
\end{widetext}Eq. (\ref{prob distribution}) now instructs us to
set the joint probability for the inputs, $i,\, j,\, k=1,\,2$, as
follows \begin{equation}
\varrho_{ijk}=\frac{1}{9} \bigl( 1+\delta_{i,\,2}\delta_{j,\,2}\delta_{k,\,2}\bigr) \,.\label{game 2 jpd}\end{equation}
While from eq. (\ref{winning cond general}) we have that given $i=j=k=1$
outputs satisfying $o_{A}^{(1)}+o_{B}^{(1)}+o_{C}^{(1)}=0$ must be
returned, given $i=j=k-1=1$ outputs satisfying $o_{A}^{(1)}+o_{B}^{(1)}+o_{C}^{(2)}\neq2$
must be returned, etc. The maximum classical total winning probability is then $\simeq 66.7\%$. See eq. (\ref{max classical}).

From \cite{Acin} we numerically have that $S_{Q}^{max}\simeq 7.37$. The maximum quantum
total winning probability is therefore $\simeq 81.9\%$, resulting in a $\simeq 15.2\%$
quantum advantage.

\section{Conclusion}

To conclude, we have presented a systematic method for the bilateral
conversion of any of the Bell inequalities, eq. (\ref{WIC}), into nonlocal games. 
In particular,
previously introduced nonlocal games are all seen to share a common
thread in this unified approach. The method is not applicable to Bell
inequalities which cannot assume a form as in eq. (\ref{WIC}), because for each of these at least one of the measurement
settings admits unequal nonvanishing coefficients $w_{M}$, eq. (\ref{Bell general}). 
This of course does not mean that another method cannot be devised
to convert any Bell inequality into a nonlocal game. However, it seems
very likely that such an increase in generality must come at the expense
of the one to one property of the mapping between the two; a nonlocal
game would then no longer fully encapsulate the unique character of
the originating Bell inequality \cite{foot3}.

In this context the work of Brukner \textit{et al}., who showed that to every Bell inequality there corresponds a communication complexity problem (CCP) \cite{Brukner}, should be mentioned (see also \cite{Pawlowski}). Indeed, any nonlocal game can be cast as a CCP. Nevertheless, no conflict arises with our previous conclusion, as not every CCP can be cast as a nonlocal game (as defined in \cite{Cleve}).

Recently, it has been argued that 
nonlocal games may be used to devise loop-hole free experimental tests
of local realism \cite{Brassard,Vaidman necklace}. To this end,
we hope that our method may prove useful. Moreover so, if
there is indeed a price to be paid for generality.

\begin{acknowledgments}
We thank N. Erez, B. Reznik and especially L. Vaidman for useful 
 discussions. We also thank M. \.{Z}ukowski for bringing to our attention ref. [17]. This work was supported 
by the Israeli Science Foundation (Grants
784-06 and 990-06).
\end{acknowledgments}


\begin{thebibliography}{10}
\bibitem{foot1}For the purpose at hand it is enough to take into
account only deterministic strategies, i.e. the output is a single
valued function of the input, since nondeterministic strategies can
at most equal the total winning probability of optimal deterministic
strategies. The averaging is thus redundant, and it suffices to consider
the ``originating'' algebraic relations, eq. (\ref{algebraic generalize}).
\bibitem{foot2}For deterministic strategies the $P\bigl(\prod_{i=1}^{n}o_{i}^{(s_{i})}=\ell_{s}\bigr)$
either vanish or equal unity. Each game's set of total winning probabilities
are then discrete instead of continuous.
\bibitem{foot3}Alternately, the one to one property can be maintained by broadening the definition of nonlocal
games to include different truth tables assigned probabilistically according to the $w_{M}$. However,  if the inequality contains
 at least
one pair of nonidentical nonvanishing $w_{M}$ corresponding to the same measurement setting, then $P_{Q}^{max}<  S_{Q}^{max}/\sum_{M} w_{M}$ (the same holds for the classical maximum winning probability as well); the result being that the full character of
the originating Bell inequality is not reflected.
\end{thebibliography}
\end{document}